\documentclass[twoside,12pt]{article}

\usepackage[sc]{mathpazo} 
\usepackage[T1]{fontenc} 
\usepackage[tmargin=25mm,bmargin=25mm,lmargin=25mm,rmargin=25mm,columnsep=20pt]{geometry} 
\usepackage{multicol} 
\usepackage[hang, small,labelfont=bf,up,textfont=it,up]{caption} 
\usepackage{booktabs} 
\usepackage{float} 
\usepackage{hyperref} 

\usepackage{graphicx}		
\usepackage{subfigure}
\usepackage{changes}			

\usepackage[semicolon,authoryear]{natbib}		

\usepackage{siunitx}			
\usepackage{textcomp}		
\usepackage{tabularx}			
\usepackage{sidecap}		
\usepackage{array}			
\usepackage{authblk}		

\usepackage{color}			
\usepackage{multirow}		
\usepackage{lmodern}	

\newcolumntype{R}{>{\raggedleft\arraybackslash}X}		
\usepackage{paralist} 

\usepackage{titlesec}

\usepackage{fancyhdr} 
\title{\vspace{-15mm}\fontsize{24pt}{10pt}\selectfont\textbf{In-situ Atom Probe Deintercalation of Lithium-Manganese-Oxide}} 

\author[1]{Björn Pfeiffer, Johannes Maier, Jonas Arlt, and Carsten Nowak}

\date{}

\begin{document}

\maketitle 

\thispagestyle{fancy}

\tableofcontents


\section{Abstract}
\label{sec:Abstract}

Atom probe tomography is routinely used for the characterisation of materials microstructures usually assuming that the microstructure is unaltered by the analysis. When analysing ionic conductors, however, gradients in the chemical potential and the electric field penetrating dielectric atom probe specimens can cause significant ionic mobility. While ionic mobility is undesired when aiming for materials characterisation it offers a strategy to manipulate materials directly in-situ in the atom probe.
Here, we present experimental results on the analysis of the ionic conductor Lithium-Manganese-Oxide with different atom probe techniques. We demonstrate that at a temperature of \SI{30}{\kelvin} characterisation of the materials microstructure is possible without measurable Li mobility. Contrary, we show that at \SI{298}{\kelvin} the material can be deintercalated in-situ in the atom probe without changing the Manganese-Oxide host structure. Combining in-situ atom probe deintercalation and subsequent conventional characterisation we demonstrate a new methodological approach to study ionic conductors even in early stages of deintercalation.


\section{Keywords}
\label{sec:Keywords}

Atom Probe, Ionic Conductor, In-situ Deintercalation, Lithium-Manganese-Oxide, Microstructure


\section{Introduction}
\label{sec:Introduction}

Atom probe tomography is routinely used for 3D chemical mapping of materials microstructures. The material is analysed by subsequent field evaporation of individual atoms or molecules from the surface. This way, the subjacent material is exposed to the surface, so that it is accessible for field evaporation, too. Thereby, the 3D information on the materials microstructure is reconstructed from a series of surface states, and although surface reconstruction effects are expected, the spatial resolution often is high enough to reveal lattice planes in crystalline materials. Thus, it is generally assumed that the materials microstructure is not significantly changed during the analysis and can be accessed using appropriate algorithms for the 3D reconstruction of the analysed volume \citep{Bas1995}.\\
However, if not all species in a multicomponent system are field evaporated and there is significant mobility of some species in the material, the chemical composition can change during analysis. This is particularly relevant for materials containing highly mobile species like Hydrogen \citep{Kesten2002} and for ionic conductors \citep{Escher2006}. Since ionic conductors are usually oxides or semiconductors, mobile species can not only be driven by gradients in the chemical potential or mechanical stress fields, but also by electric fields penetrating dielectric atom probe specimens \citep{Silaeva2013,Silaeva2014,Greiwe2014}. Therefore, the influence of the electric field between specimen and counter electrode in an atom probe has to be considered. While mobility of species during atom probe analysis is undesired when aiming for microstructural characterisation (cf. \cite{Schmitz2010,Santhanagopalan2015,Devaraj2015,Maier2016}) it offers a strategy for controlled specimen manipulation in an atom probe.\\
Here, we present enhanced atom probe analyses of Lithium-Manganese-Oxide (LiMn$_{2}$O$_{4}$; short: LMO) as model system for an ionic conducting material relevant for electrochemical energy conversion \citep{Zhang2015}. In the initial state the LMO has a cubic spinel structure resulting in an isotropic diffusion network for the mobile Li-ions. This Li mobility is central for the atom probe analyses presented here. Adjusting specimen base temperature and electric field for field evaporation enables operating the atom probe in different modes. We show that at \SI{30}{\kelvin} and high electric fields conventional microstructural analysis of specimens is possible without measurable Li mobility. This is particularly important since this aspect is not addressed in the literature on atom probe analysis of Lithium-(Nickel)-Manganese-Oxide so far \citep{Devaraj2015,Maier2016}. In contrast, for a temperature of \SI{298}{\kelvin} and lower electric fields we demonstrate that in-situ deintercalation experiments can be performed in the atom probe without changing the Mn-O host structure. By combining both modes we show that during deintercalation of LMO complex microstructural features develop and discuss their impact on Li transport in LMO.


\section{Materials and Methods}
\label{sec:MaterialsandMethods}

\subsection{Lithium-Manganese-Oxide}
\label{subsec:LMO}

At room temperature Lithium-Manganese-Oxide with the stoichiometry LiMn$_{2}$O$_{4}$ has a cubic spinel structure (space group $Fd\overline{3}m$) with lattice parameter $a \approx$ \SI{8.24}{\angstrom}. The O atoms are located at 32e positions, Li at tetrahedral 8a, and Mn at octahedral 16d sites \citep{Shimakawa1997, Rodriguez-Carvajal1998, Huang2011}. Deintercalating Li$_{x}$Mn$_{2}$O$_{4}$ in the range $0.5<x<1$ leads to a homogeneous single phase material with lattice parameter decreasing to approximately \SI{8.15}{\angstrom} for $x=0.5$ \citep{Ohzuku1990, Kanamura1996}. At $x \approx 0.5$ Li ordering in the cubic spinel structure takes place \citep{Ohzuku1990, Kucza1999, Zhang2015}. For $0<x<0.5$ two phases are formed: a Li-rich phase with $x \approx 0.5$ and a cubic spinel-related phase of MnO$_{2}$ with small solubility of Li. Deintercalating LiMn$_{2}$O$_{4}$ all occurring phases exhibit the space group $Fd\overline{3}m$ \citep{Ohzuku1990}.\\
In \citet{Li2000} the deintercalated system is additionally observed in a non-equilibrium state with two-phase regions at around $0.25<x<0.55$ and $0.60<x<0.95$. After several days equilibrium is reached and the phases reported in \citet{Ohzuku1990} are observed. Since structural information on Li$_{x}$Mn$_{2}$O$_{4}$ was mostly obtained using X-ray diffraction it represents averages over relatively large sample volumes and gives only indirect information on the local distribution of the occurring phases in the material.\\[2mm]
For the diffusion coefficient $D_{Li}(x)$ of Li in Li$_{x}$Mn$_{2}$O$_{4}$ values in the range from \SI{E-11}{\centi\meter\squared\per\second} to \SI{E-8}{\centi\meter\squared\per\second} are reported with qualitatively different and even contradictory dependencies on the Li content $x$. \cite{Yang1999} showed a decreasing diffusion coefficient with increasing $x$, whereas \cite{Bach1998} measured a maximum for the diffusion coefficient at $x \approx 0.55$. Two maxima at $x \approx 0.3$ and $x \approx 0.7$ for $D_{Li}(x)$ and similarly for the conductivity are reported by \cite{Ouyang2004a}. The reason for these differences are not clear. Although \cite{Bach1998} discuss a possible influence of the preparation method of LiMn$_{2}$O$_{4}$ on the Li mobility, a detailed study on the influences of defects and the microstructure on the mobility of Li in LiMn$_{2}$O$_{4}$ is still missing.\\[2mm]
For specimen preparation Lithium-Manganese(III,IV)-Oxide (LiMn$_{2}$O$_{4}$) from Sigma Aldrich with a particle size $<$\SI{5}{\micro\meter} is used. The facetted particles are expected to be monocrystalline and transmission electron microscopy (TEM) diffraction analysis after specimen preparation reveals single crystallinity for the specimens analysed in this work. Selected area diffraction (SAD) patterns of the specimen apex regions fit to the expected cubic spinel structure with a lattice parameter of around \SI{8.24}{\angstrom}.\\

\subsection{Experimental}
\label{subsec:Experimental}

Specimen preparation was carried out using a FEI Nova NanoLab 600 focused ion beam (FIB) with integrated scanning electron microscope (SEM). Lift out of the LiMn$_{2}$O$_{4}$ particles and the following connection to a tungsten support tip was done with a micromanipulator and deposition of standard platinum precursor trimethyl(methylcyclopentadienyl)platinum(IV). The specimens were sharpened by side cuts of the Ga-ion beam at \SI{30}{\kilo\electronvolt} beam energy to a full shank angle of around \SI{12}{\degree} or by annular milling \citep{Larson1999}. The last step of preparation was irradiation of the specimen in axial direction with the ion beam at \SI{5}{\kilo\electronvolt} beam energy to reduce the volume implanted with Ga \citep{Miller2007a}. This way, apex radii of as-prepared specimens between \SI{10}{\nano\meter} and \SI{30}{\nano\meter} were achieved. The morphology and crystal structure of the LMO specimens before and after atom probe analysis was investigated with a Philips CM12 or CM30 TEM using a single tilt holder.\\[2mm]
Laser-assisted atom probe analysis was performed with an in house constructed system described in \cite{Maier2016}. The specimen base temperature between \SI{30}{\kelvin} and \SI{298}{\kelvin} was controlled by a combination of closed cycle He cryostat and resistive heating. The background pressure was $\SI{6E-8}{\pascal}$ at \SI{298}{\kelvin} and $\SI{1E-8}{\pascal}$ at \SI{30}{\kelvin}, respectively. The laser with a pulse duration of \SI{15}{\pico\second}, a focus diameter of \SI{125}{\micro\meter} at the specimen and a wavelength of \SI{355}{\nano\meter} was operated at a repetition rate of \SI{200}{\kilo\hertz}. Depending on the mode of analysis pulse energies were varied from \SI{25}{\nano\joule} to \SI{200}{\nano\joule}.\\[2mm]
For this study two different modes of atom probe analysis were used. Conventional atom probe analysis was performed at \SI{27.5}{\nano\joule} pulse energy with a specimen base temperature of \SI{30}{\kelvin} and automated voltage adjustment to achieve a detection rate in the range of 0.004 - 0.01 ions per pulse corresponding to \SI{8E2}{\per\second} to \SI{2E3}{\per\second}. For the in-situ deintercalation experiments first a conventional analysis up to a specific voltage $V_{max}$ was performed to prepare a well-defined initial specimen state. Then the voltage was set to a constant value $V$ in the range of \SI{50}{\percent} up to \SI{80}{\percent} of $V_{max}$, the sample base temperature was increased up to \SI{298}{\kelvin} and a higher pulse energy in the range of \SI{70}{\nano\joule} to \SI{200}{\nano\joule} was used yielding a rate of detected Li-ions in the range of \SI{1E3}{\per\second} to \SI{1E5}{\per\second} after a short time. The detection rate exhibits strong temporal fluctuations and decreases down to $\approx \SI{30}{\per\second}$ after several hours (cf. section \ref{subsec:APT298K}). Since the electric field strength at the apex is reduced compared to conventional atom probe analysis by setting a constant voltage $V<V_{max}$, the low evaporation field for Li enables field evaporation of solely Li without affecting the Mn-O host structure (see section \ref{subsec:APT298K}).


\section{Results and Discussion}
\label{sec:ResultsDiscussion}

\subsection{Lithium Mobility under Conventional Atom Probe Analysis at 30K}
\label{subsec:APT30K}

\begin{figure}[H]
  \centering  \includegraphics[width=0.95\linewidth]{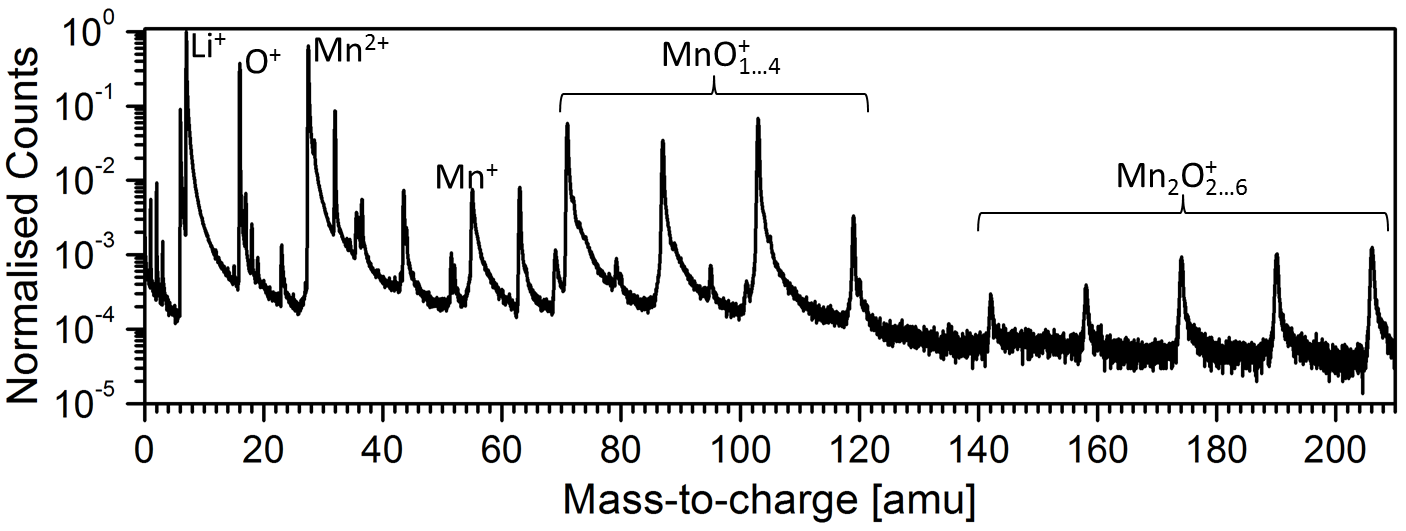}
  \caption[]{Logarithmically plotted normalised mass spectrum of an atom probe analysis of LMO at 30$\,$K. Only major peaks are labelled. The spectrum is truncated at 210$\,$amu as no further peaks are observed for higher mass-to-charge ratios.}
  \label{fig:Figure1}
	\end{figure}

\begin{figure}[H]
  \centering  \includegraphics[width=0.8\linewidth]{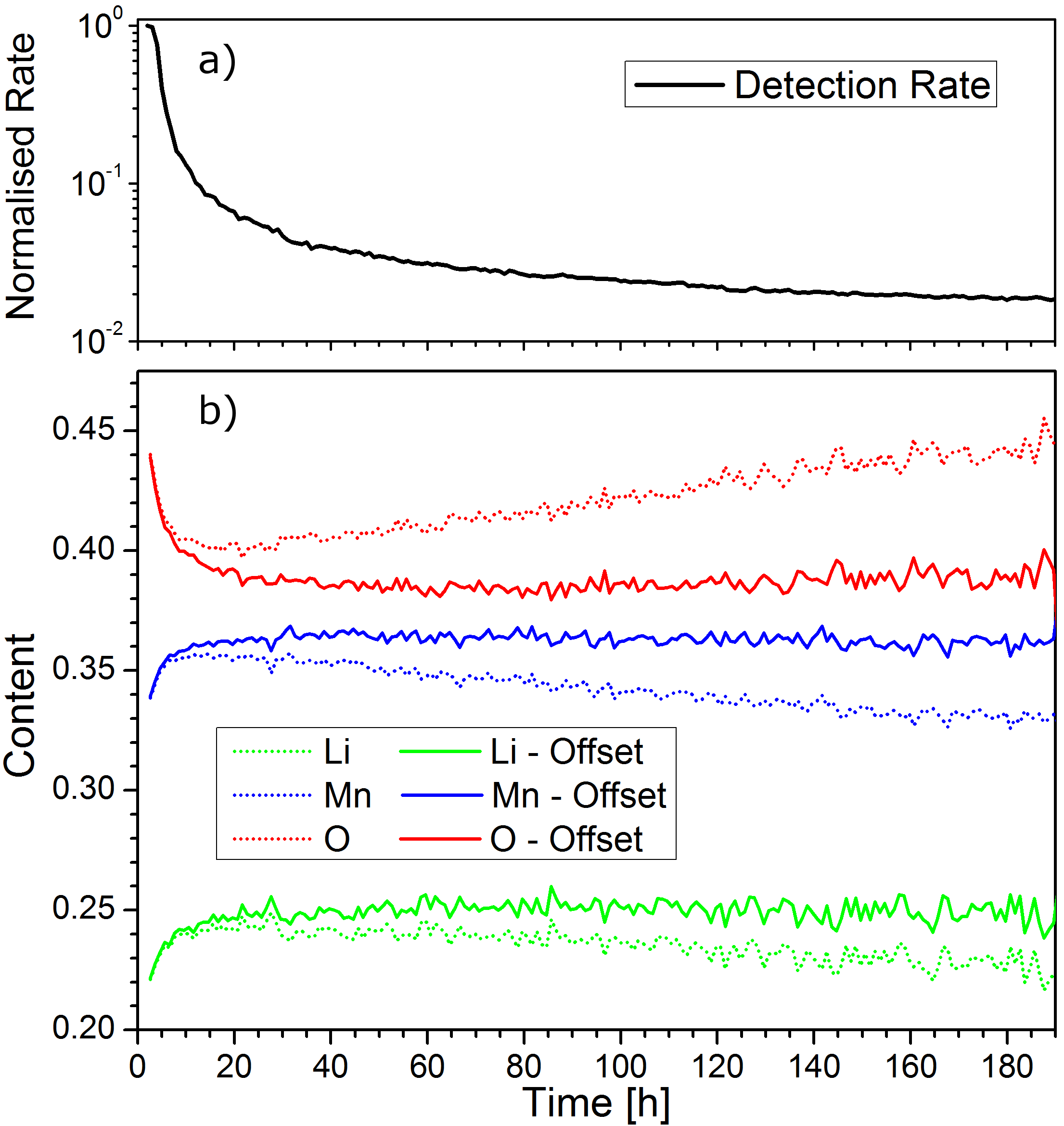}
  \caption[]{a) Logarithmically plotted normalised detection rate for an atom probe analysis at a constant base voltage of 5.8$\,$kV and a laser pulse energy of 27.5$\,$nJ. The automatic voltage adjustment of the preceding conventional atom probe analysis was stopped at $t=0$. The maximum detection rate prior to normalisation was 3.5 million per hour. b) Content of Li, Mn or O for the detected ions displayed in a) plotted with dotted lines. The solid lines show the content after subtraction of a constant offset of 4500 detected O atoms per hour to consider O supply from the residual gas.}
  \label{fig:Figure2}
	\end{figure}

A typical mass spectrum for a conventional atom probe analysis of LMO at \SI{30}{\kelvin} is shown in Fig. \ref{fig:Figure1}. Besides the elemental peaks for Li$^{+}$, Mn$^{+}$, Mn$^{2+}$, O$^{+}$, O$_{2}^{+}$ various Mn$_{x}$O$_{y}^{z+}$ peaks are observed. Additional peaks, not directly related to the analysed material, are those of Hydrogen and Gallium resulting from the residual gas in the UHV chamber and the FIB based specimen preparation, respectively. The peak at \SI{71}{amu}, assigned to MnO$^{+}$ also covers the events from the minor isotope $^{71}$Ga$^{+}$, but since the signal for $^{69}$Ga$^{+}$ is relatively low, this overlap is of minor relevance and not expected to influence the results crucially. By calculating the atomic ratios a stoichiometry of approximately Li$_{1.2}$Mn$_{2}$O$_{2.3}$ is obtained. This shows that O is detected substoichiometrically, which is a known effect in atom probe tomography of oxides (e.g. \cite{Karahka2015}). Nevertheless the data shows a uniform distribution of all elements and the spatial resolution is high enough to reveal lattice planes in the reconstruction \citep{Maier2016}.\\
To measure eventual Li mobility at \SI{30}{\kelvin} the automatic base voltage adjustment was stopped during a conventional atom probe analysis at a base voltage of \SI{5.8}{\kilo\volt}. The temporal evolution of the resulting normalised detection rate is displayed in Fig. \ref{fig:Figure2} a). The dotted lines in Fig. \ref{fig:Figure2} b) illustrate the average content of Li, Mn, and O. These graphs show a different behaviour before and after a characteristic time of \SI{20}{\hour}. Initially, the curve for O is decreasing while the curves for Li and Mn are increasing. From \SI{20}{\hour} on the curve for O is increasing, while the curves for Li and Mn are decreasing.\\
The decreasing detection rate in Fig. \ref{fig:Figure2} a) is naturally explained because further field evaporation causes tip blunting and thus a decreasing electric field strength at the specimen apex. However, if Li-ions were mobile in the specimen under the given conditions the decreasing detection rate would be only expected for the immobile host species Mn and O. For Li a net flux in the direction of the apex would be expected, either driven by the electric field inside the specimen or by a gradient in the chemical potential. Since elemental Li has a low theoretical field evaporation field strength of \SI{14}{\volt\per\nano\meter} compared to $\SI{30}{\volt\per\nano\meter}$ for Mn \citep[p. 492; image hump model]{Miller1996} and the experimental value of $\approx \SI{28}{\volt\per\nano\meter}$ for LMO, Li field evaporation at the apex and thus an increasing Li content would be expected with time.\\[2mm]
The temporal evolution of the Li content is analysed based on the data displayed in Fig. \ref{fig:Figure2} b). For the content of Li, Mn, and O two different regimes can be observed. In the first \SI{20}{\hour} the dominant effect is the decreasing electric field strength. Because the detection rate for O$^{+}$ and O$_{2}^{+}$ is observed to decrease stronger than the one for other atomic or molecular species, the average content of O in this region becomes lower, in accordance with the results of \citet{Devaraj2013}. For the same reason, the average content of Li and Mn increases. From \SI{20}{\hour} on this effect plays a minor role and the average content of O increases, whereas the average content of Li and Mn decreases (dotted lines in Fig. \ref{fig:Figure2}).\\
If there were significant Li transport in the sample, the Li content would be expected to increase, thus, the observed decrease cannot be naturally explained this way. On the other hand, the observed increase in the O content is not attributed to mobile O in the sample since O has no significant mobility at $\SI{30}{\kelvin}$ under field free conditions and the electrostatic field would cause a driving force for O transport into the sample. Besides the sample, another source of O can be the residual gas in the atom probe and a simple estimate yields that several thousand molecules from the residual gas are impinging on the specimen apex per hour.\footnote{For a total background pressure of $\approx$ $\SI{1E-8}{\pascal}$ at \SI{30}{\kelvin}, a typical specimen apex area of around \SI{5E3}{\nano\meter\squared} and the molecular mass of O$_{2}$ or H$_{2}$O the Hertz-Knudsen equation gives approximately \SI{2E4}{Molecules\per\hour} impinging on the specimen apex.} Since the impinging rate is expected to be influenced by the electric field in the vicinity of the sample \citep{Eekelen1970} and the reaction and ionization probability of the residual gas molecules are unknown a deterministic calculation of the contribution of residual gas molecules to the O content is not straight forward. However, subtracting a constant O supply rate from the residual gas prior to calculation of the content of Li, Mn, and O yields different temporal evolutions of the Li, Mn, and O content. Particularly, for a supply rate of 4500 O atoms per hour from the residual gas the content of Li, Mn, and O becomes constant from 20 h on as shown by the solid lines in Fig. \ref{fig:Figure2} b). Such a constant content for all three species would be expected for Li being practically immobile in the sample and evaporating only together with the Mn-O host structure. A further indication for no significant Li transport in the sample but significant O supply from the residual gas is given by the average number of Li, Mn, and O atoms present in the detected ions when omitting noise. Not considering O supply from the residual gas, the average composition of these ions is Li$_{0.34}$Mn$_{0.50}$O$_{0.55-0.65}$ with an increasing O content with time. Subtracting a constant supply rate of 4500 O atoms per hour from the residual gas the composition becomes Li$_{0.34}$Mn$_{0.50}$O$_{0.52}$, independent of time for times from 20 h on. In both cases, the Li content is observed to be constant, indicating no Li transport in the sample.\\
Thus, it can be concluded that under conventional atom probe analysis conditions at \SI{30}{\kelvin} no transport of Li, Mn, and O is observed. From the noise level of the atom probe analysis a possible Li flux can be estimated to be below $j_{Li} <\,$\SI{1E9}{\per\centi\meter\squared\per\second}.
	
\subsection{In-situ Atom Probe Deintercalation at 298K}
\label{subsec:APT298K}

\begin{figure}[H]
  \centering  \includegraphics[width=0.95\linewidth]{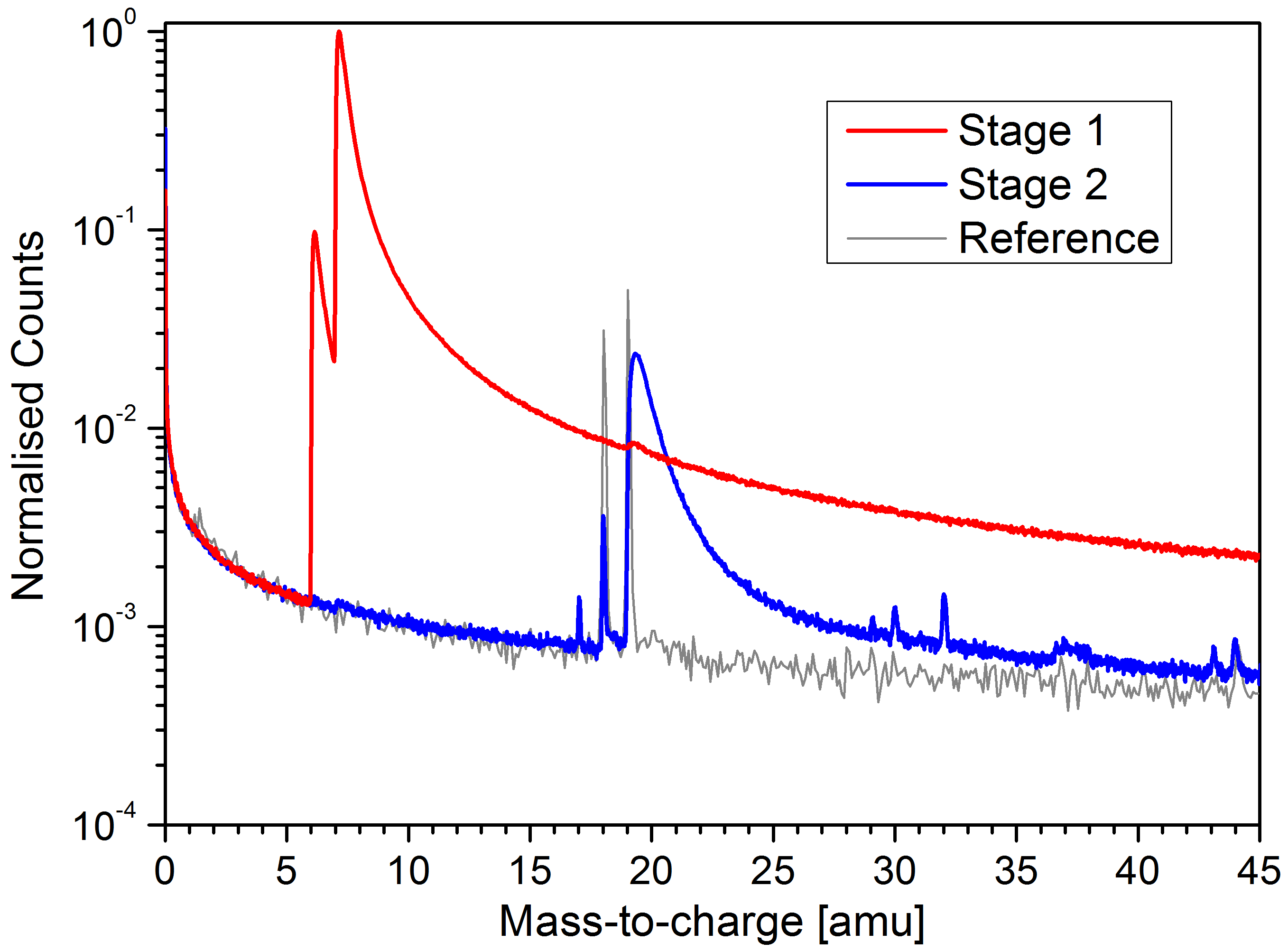}
  \caption[]{Logarithmically plotted mass spectra for the two stages of in-situ deintercalation of LMO at 298$\,$K (red and blue curve). The grey curve shows a mass spectrum for a tungsten sample under similar field evaporation conditions. All spectra are normalised to the same noise level for mass-to-charge ratios lower than 5.}
  \label{fig:Figure3}
	\end{figure}

\begin{figure}[H]
  \centering  \includegraphics[width=0.4\linewidth]{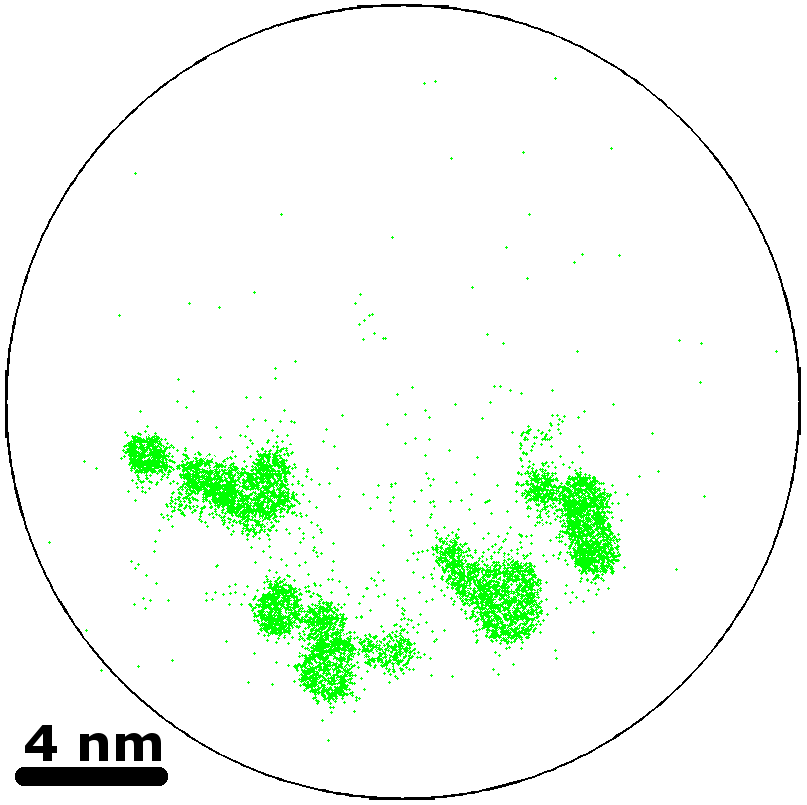}
  \caption[]{Spatial distribution of the detected ions during stage 1 of the in-situ deintercalation reaction of LMO at 298$\,$K. Displayed are 50.000 ions assigned to the elemental Li peaks. The circle indicates the active area of the detector in the atom probe, the scale represents the corresponding length on the specimen.}
  \label{fig:Figure4}
	\end{figure}

The mass spectra for the in-situ deintercalation experiments differ significantly from those for the conventional atom probe analysis displayed in Fig. \ref{fig:Figure1}. Two characteristic types of mass spectra are observed which usually appear subsequently and are attributed to two consecutive characteristic stages of deintercalation as discussed below. Fig. \ref{fig:Figure3} shows the normalised mass spectra.\\
For stage 1 the mass spectrum is formed by the peaks for the two Li isotopes at \SI{6}{amu} and \SI{7}{amu}. Besides an additional small peak at $\SI{19.3}{amu}$ no further peaks are observed. Particularly no peaks for Mn$^{+}$, Mn$^{2+}$, O$^{+}$, O$_{2}^{+}$ or Mn$_{x}$O$_{y}^{z+}$ occur, revealing that the Mn-O host structure is not subject to field evaporation during in-situ deintercalation at \SI{298}K.\\
For stage 2 the mass spectrum exhibits no peaks for Li at \SI{6}{amu} and \SI{7}{amu}. Instead, the spectrum is dominated by a peak at $\SI{19.3}{amu}$. Additionally, small peaks at $\SI{17}{amu}$ (OH$^+$), $\SI{18}{amu}$ (OH$_{2}^{+}$), and $\SI{32}{amu}$ (O$_{2}^{+}$) are observed as well as peaks at \SI{29}{amu}, \SI{30}{amu}, \SI{43}{amu}, \SI{44}{amu}, and a broad peak around \SI{36.5}{amu}. To assign the peak at $\SI{19.3}{amu}$, control experiments with a tungsten sample under similar field evaporation conditions were performed, yielding the reference mass spectrum in Fig. \ref{fig:Figure3} with peaks at $\SI{17}{amu}$, $\SI{18}{amu}$, $\SI{19}{amu}$, and $\SI{44}{amu}$. Compared to the mass spectrum of stage 2, the peaks at $\SI{19}{amu}$ and $\SI{19.3}{amu}$ differ in both, peak position and peak shape. Possible assignments for the peak at $\SI{19.3}{amu}$ are OH$_{3}^{+}$, MnH$_{3}^{3+}$, and $^{7}$Li(OH)$_{3}^{3+}$ (cf. Tab. \ref{tab:Table1}). In the case of OH$_{3}^{+}$, the observed peak shape could eventually be explained by delayed emission due to inhomogenous optical absorption of the sample \citep{Vella2011,Kelly2014}. However, as the other peaks in the mass spectrum for stage 2 exhibit no peak broadening this explanation seems unlikely. For an assignment of MnH$_{3}^{3+}$ a lack of field evaporated Li and O would make this assignment unlikely as well. Thus, the peak at $\SI{19.3}{amu}$ is attributed to $^{7}$Li(OH)$_{3}^{3+}$, although the presence of such a high ionization state is unexpected under the electric field conditions present during stage 2. The assignment is supported by the spatial distribution of Li in the reconstructed volume as discussed in section \ref{subsec:MspD}. Since the field evaporation rate for stage 2 is comparable with the impinging rate of residual gas molecules onto the specimen (cf. section \ref{subsec:APT30K}), the O and H necessary to form the Li(OH)$_{3}$ molecule is expected to originate from the residual gas. Due to the peak broadening, the $^{6}$Li(OH)$_{3}^{3+}$ peak expected at $\SI{19}{amu}$ is not resolved. Considering this, there is deintercalation in stage 2, too, where the Li from the LMO reacts with residual gas and is detected as lithium hydroxide. As in stage 1, the Mn-O host structure is not or only marginally field evaporated.\\
\begin{table}[H]
\centering
\begin{tabularx}{0.48\linewidth}{Xr}
Mass [amu] & Ions \\
\midrule
17 & OH$^{+}$ \\
18 & OH$_{2}^{+}$ \\
$\approx$ 19 & $^{7}$Li(OH)$_{3}^{3+}$, MnH$_{3}^{3+}$, OH$_{3}^{+}$ \\
29 & $^{6}$Li$^{7}$LiO$^{+}$, $^{7}$Li(OH)$_{3}^{2+}$ \\
30 & $^{7}$Li$_{2}$O$^{+}$ \\
32 & O$_{2}^{+}$ \\
$\approx$ 36.5 & $^{7}$Li$_{3}$OH$_{x}^{+}$, MnOH$_{x}^{2+}$, O$_{2}$H$_{5}^{+}$ \\
43 & $^{7}$Li(OH$_{2}$)$_{2}^{+}$ \\
44 & CO$_{2}^{+}$ \\
\bottomrule
\end{tabularx}
\caption{Possible peak assignment\added{s} for the mass spectrum shown in Fig. \ref{fig:Figure3} for stage 2 of the in-situ deintercalation of LMO.}
\label{tab:Table1}
\end{table}
The spatial distribution of the detected Li-ions for stage 1 is displayed in Fig. \ref{fig:Figure4}. The Li signal is observed to be inhomogeneously distributed with Li-ions appearing at about 10 spots. These spots have a characteristic lifetime from several seconds up to several minutes, then the individual spots disappear and new spots appear at new or previous positions. The probability for a spot to occur is not equally distributed over the sample surface, typically only a part of the sample surface, often one side, exhibits spots while in large areas no spots are observed during the course of an experiment. The temporally fluctuating spots yield detection rates up to $10^5$ counts per second with an average in the range of $10^3\,$s$^{-1}$. The total amount of field evaporated ions during stage 1 varies between some tens of million up to several hundreds of million Li-ions. In stage 2 the field evaporation is spatially and temporally homogeneous (not shown) with a field evaporation rate of around $30\,$s$^{-1}$ slightly above the noise level. The relation between the characteristic field evaporation behaviour of the two stages and the specimens microstructure is discussed in the following section.\\
Since the distribution of ions at the $\SI{19.3}{amu}$ peak is spatially homogeneous for both stages, it is likely that the field evaporation behaviour of stage 2 is present during stage 1, too, and the small peaks of stage 2 in Fig. \ref{fig:Figure3} are covered under the tail of the several magnitudes higher peaks for elemental Li.

\subsection{Microstructure Subsequent to Deintercalation}
\label{subsec:MspD}

\begin{figure}[H]
  \centering  \includegraphics[width=0.9\linewidth]{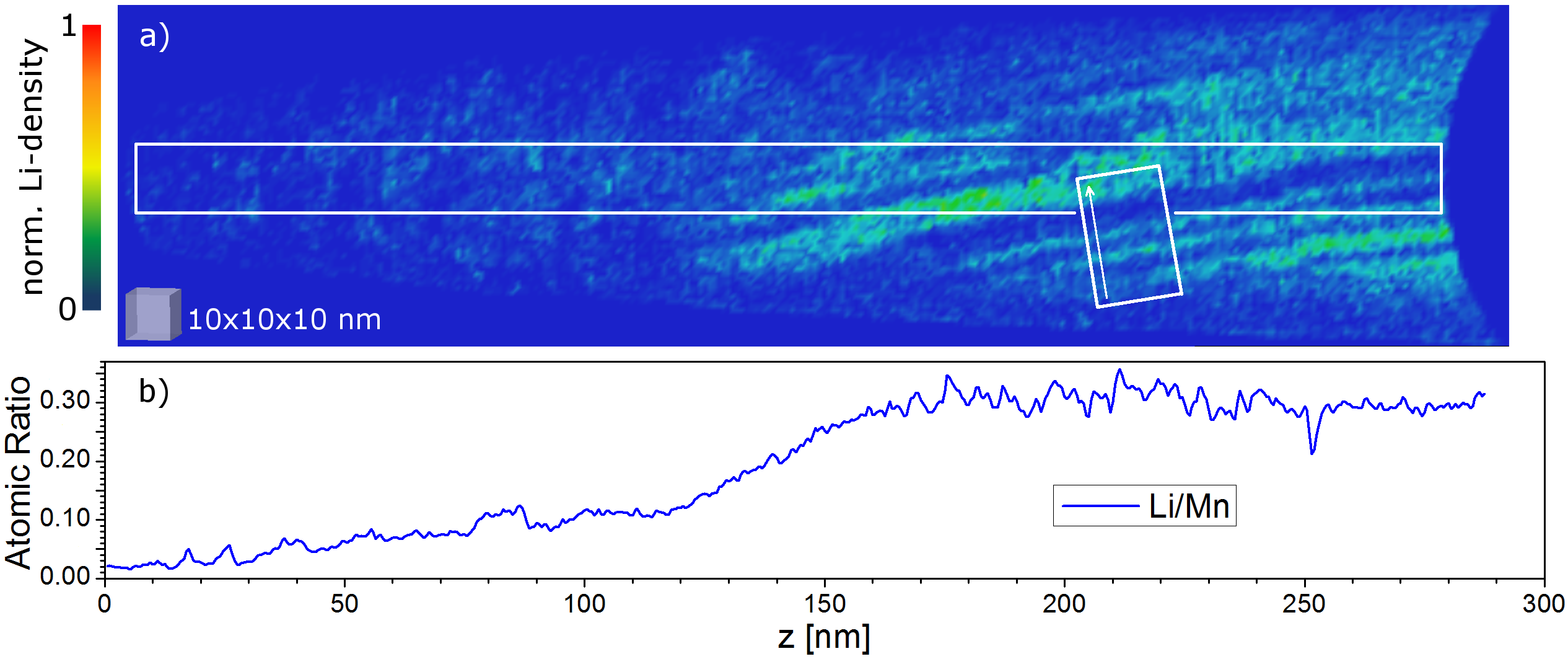}
  \caption[]{a) Li density in a 1$\,$nm thick slice through the reconstruction of a L$_{x}$MO specimen subsequent to in-situ deintercalation for $\approx$24$\,$h at stage 1 and $\approx$120$\,$h at stage 2. The white rectangle with the arrow shows the region analysed in Fig. \ref{fig:Figure6}. b) Ratio of Li to Mn (complex ions split) calculated for the cylindrical volume with 15$\,$nm diameter in the center of the reconstruction marked with a white rectangle in a).}
  \label{fig:Figure5}
	\end{figure}

Subsequent to deintercalation the microstructure of L$_{x}$MO (for Li$_{x}$Mn$_{2}$O$_{4}$ with $x<1$) was characterised with conventional atom probe analysis at 30 K as described in \ref{subsec:Experimental}. For all specimens an inhomogeneous Li distribution with alternating Li-rich and Li-depleted regions was found. Fig. \ref{fig:Figure5} a) shows the Li density in a slice of the reconstruction of a L$_{x}$MO specimen from which $\approx 10^{9}$ Li ions were deintercalated. During this deintercalation stage 1 was active for $\approx$\SI{24}{\hour} and stage 2 for $\approx$\SI{120}{\hour}. In the region from the apex up to a reconstruction depth of around \SI{120}{\nano\meter} no specific structures are observed in the Li density. For a reconstruction depth of more than \SI{120}{\nano\meter} alternating Li-rich and Li-depleted regions are found. Although lattice planes are found in the reconstruction, the crystallographic orientation of the regions of different Li density cannot be determined unambiguously.\\
The absolute ratio of Li to Mn along the centre of the reconstruction of Fig. \ref{fig:Figure5} a) is displayed in Fig. \ref{fig:Figure5} b). For the first \SI{120}{\nano\meter} besides some fluctuations a continuously increasing ratio of Li to Mn from 0.02 to 0.12 is observed. After a transition region from \SI{120}{\nano\meter} to \SI{160}{\nano\meter} with an increased slope the ratio of Li to Mn exhibits no dependence on the reconstruction depth and appears constant with some fluctuations.\\	
	\begin{figure}[H]
  \centering  \includegraphics[width=0.8\linewidth]{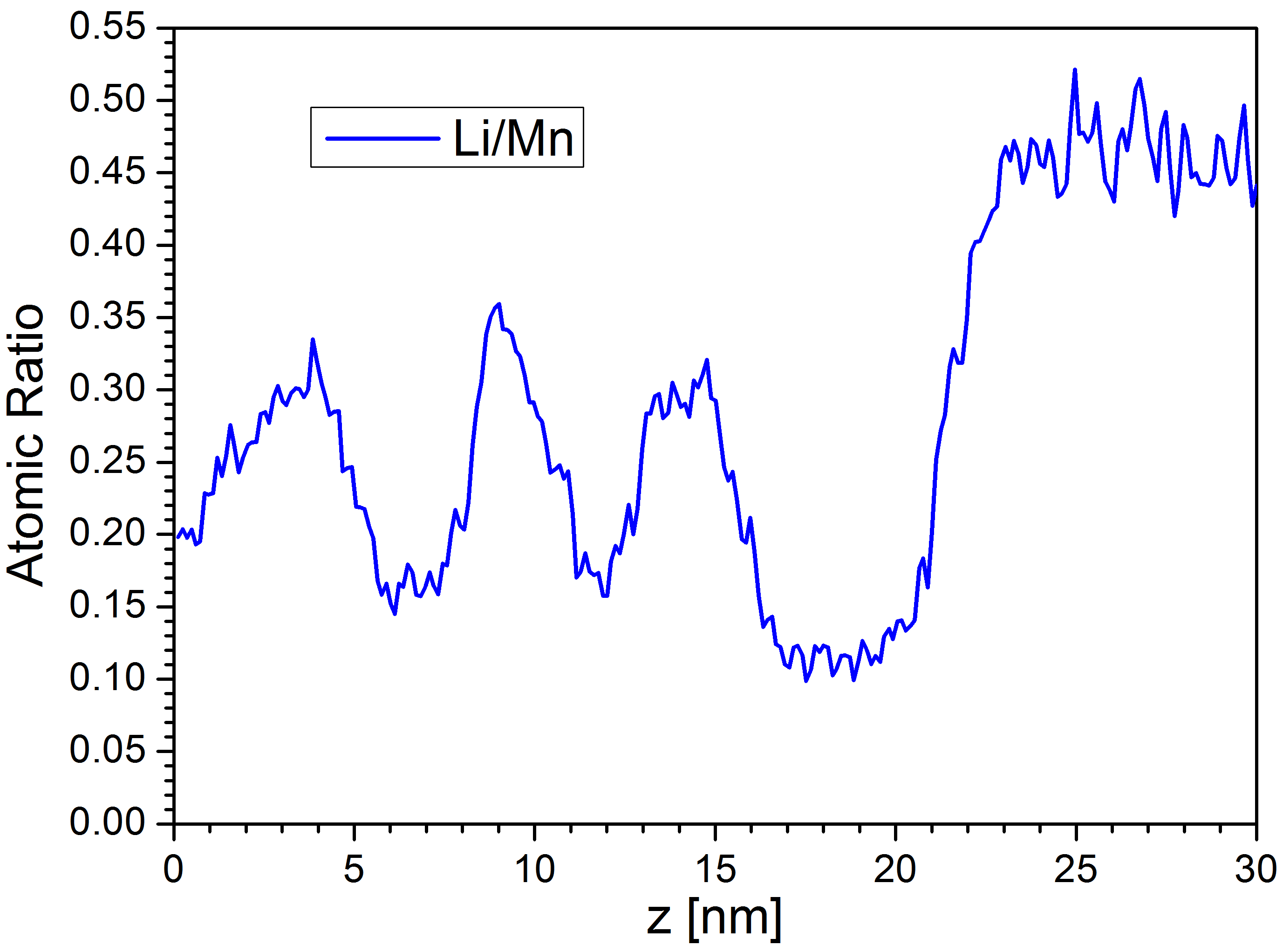}
  \caption[]{Ratio of Li to Mn (complex ions split) perpendicular to the alternating regions of different Li density for the part of the reconstruction marked in Fig. \ref{fig:Figure5} a).}
  \label{fig:Figure6}
	\end{figure}	
The ratio of Li to Mn perpendicular to the alternating regions of different Li density for the part of the reconstruction marked in Fig. \ref{fig:Figure5} a) is displayed in Fig. \ref{fig:Figure6}. The ratio is alternating between three different ranges of values with relatively sharp transitions between them. A ratio between 0.45 and 0.5 is typically found for not or just slightly deintercalated L$_{x}$MO specimens (cf. \cite{Maier2016}). The other ratios from 0.35 to 0.3 and from 0.15 to 0.1 do not fit to the equilibrium phases of Lithium-Manganese-Oxide reported in \cite{Ohzuku1990}. However, they are similar to those reported for the two non-equilibrium phases observed in chemical lithiation experiments of $\lambda$-MnO$_2$ \citep{Li2000}. As similar compositions were observed for all deintercalated specimens, this indicates that the in-situ atom probe deintercalation yields an early non-equilibrium stage of deintercalated L$_{x}$MO.\\[2mm]
A microscopic model for the deintercalation mechanism can be developed by relating the two regions in Fig. \ref{fig:Figure5} a) to the two stages of deintercalation introduced in section \ref{subsec:APT298K}: the first part in Fig. \ref{fig:Figure5} up to a depth of $\SI{120}{\nano\meter}$ can be associated with the homogeneous field evaporation behaviour at stage 2. The spatially and temporally homogeneous field evaporation and the absence of specific structures in the reconstruction indicate temporally and spatially homogeneous volume diffusion of Li as deintercalation mechanism. Furthermore, comparing the Li content in the first $\SI{120}{\nano\meter}$ of Fig. \ref{fig:Figure5} with the inhomogeneous microstructure behind shows that the amount of lacking Li approximately equals the amount of ions detected at $\SI{19.3}{amu}$ (cf. Fig. \ref{fig:Figure3}).\\
On the other hand the inhomogeneous microstructure in Fig. \ref{fig:Figure5} a) most likely is related to the inhomogeneous field evaporation behaviour of stage 1. This interpretation is supported by deintercalation experiments stopped at stage 1 where the inhomogeneous microstructure was observed to extend to the specimen apex region (not shown). Since the diffusion coefficient of Li in Lithium-Manganese-Oxide can vary up to several orders of magnitude depending on the Li content \citep{Bach1998,Yang1999,Ouyang2004a} Li diffusion is expected to be strongly localised in the inhomogeneous microstructure. However, there is no direct experimental evidence to differentiate whether Li transport predominantly occurs in one of the observed most likely metastable Li-depleted phases or along the interface of different phases.\\
As in Fig. \ref{fig:Figure5} the Li-rich and Li-depleted regions are often nearly parallel and usually oriented such that their surface normal is not perpendicular to the specimens axis of symmetry. Thus, preferential Li transport along the Li-depleted regions causes an azimuthally inhomogeneous Li distribution, in accordance with the observation that Li is typically detected on one side of the detector during stage 1 (cf. Fig. \ref{fig:Figure4}). In this process Li-ions do not necessarily have to be field evaporated once they reach the specimen surface. It is also conceivable that first surface diffusion takes place and localised field evaporation occurs at specific sites.\\[2mm]
Hence, the spatially and temporally inhomogeneous in-situ deintercalation at stage 1 is expected to be correlated to the microstructural evolution and phase transition in the L$_{x}$MO. Nevertheless, also during stage 1 the deintercalation mechanism of stage 2 is expected to be active, although at very low rates.


\section{Conclusion}
\label{sec:Conclusion}

In this work the influence of Li-ion mobility in LiMn$_{2}$O$_{4}$ on atom probe analysis of the material was investigated.
It was demonstrated that for conventional atom probe analysis at \SI{30}{\kelvin} Li-ion mobility is suppressed below the detection limit so that the materials microstructure can be characterised reliably. At \SI{298}{\kelvin} a substantial increase in mobility can be achieved, enabling in-situ deintercalation of the material in the atom probe without affecting the Mn-O host structure significantly. Two different modes of deintercalation have been observed: spatially and temporally inhomogeneous deintercalation at high rates at the beginning, followed by homogeneous deintercalation at low rates. Characterisation by conventional atom probe analysis revealed a complex non-equilibrium microstructure with alternating Li-rich and Li-depleted regions subsequent to deintercalation. A model for the deintercalation mechanism based on homogeneous volume transport
and localised transport in the Li-depleted regions was proposed.\\
The methodological approach presented here allows in-situ preparation and subsequent characterisation of non-equilibrium microstructures in an ionic conductor. This way, insights into the microstructural evolution accompanying deintercalation processes can be obtained. Particularly for nanowire Li-ion batteries new insights into the mechanisms related to electrochemical energy conversion are expected, that are currently not accessible with other in-situ techniques \citep{Lee2013,Lee2015}.


\section{Acknowledgments}
\label{sec:Acknowledgments}

Financial support by the German Research Foundation (SFB 1073, TP C05) and the Graduate program for Energy Storage and Electromobility in Lower Saxony (GEENI) is gratefully acknowledged.


\newpage

\bibliography{literature}


\end{document}